\documentclass[12pt]{article}
\usepackage{graphicx}
\usepackage{amsfonts}
\usepackage{bm}
\newcommand{\vett}[1]{\mathbf{#1}}
\newcommand{\dif}{{\rm d}}

\renewcommand{\phi}{{\varphi}}
\newcommand{\equal}{\buildrel {\rm def} \over {=} }

\title{Chemical bond by Newtonian trajectories in the $H_2^+$ ion}

\author{A. Carati\thanks{Department of Mathematics, Universit\`{a}
    degli Studi di Milano -
    E-mail: \texttt{andrea.carati@unimi.it}}
  \and
  L.Galgani\footnotemark[1]
  \addtocounter{footnote}{1}
  \and
  F. Gangemi\thanks{DMMT, Universit\`a di Brescia, Viale Europa 11,
    I-25123 Brescia, Italy}
  \and
  R. Gangemi\footnotemark[3]
}

\begin{document} 

\maketitle

\begin{abstract}
  According to the   correspondence principle,
  classical mechanics and quantum mechanics  agree in the semiclassical
  limit,  although presently it has become more and more clear how intriguing
  would be to try to  fix a boundary between them.   Here we give a
  significant example in which the agreement concerns  Newtonian trajectories
  of an electron with initial data  corresponding  to a quantum   ground state.
  The example is the simplest case in which a chemical bond occurs,
  i.e. the $H_2^+$ ion. By molecular dynamics simulations for the full system
  (two  protons and one electron) we show that   there
  exist initial data  producing an ``effective potential'' among the protons,
  which superposes in a surprisingly good way the quantum one computed in the
  Born-Oppenheimer approximation (Fig~\ref{fig:1}). Preliminarily,  following the
  perturbation procedure first  exhibited  by Born and Heisenberg in the year
  1924, we recall why an effective potential should exist in a classical frame,
  and also describe the numerical procedure employed  in computing it. 
\end{abstract}


\section{Introduction}
In condensed-matter physics one deals with models in which matter is
constituted of ions and electrons, with mutual Coulomb interactions.
Since the first works of Born and Heisenberg \cite{bornheis} and of
Born and Oppenheimer \cite{bornopp}, it was shown that a reduced
dynamics can be obtained for the motions of the ions only, in which
the attractive role of the electrons is taken into account through
suitable ``effective potentials'' among the ions.

However, such potentials, which were first shown to exist by Born and
Heisenberg in the year 1924 by the methods of classical perturbation
theory, are too complicated to be computed analytically. So,
phenomenological potentials were used, and only after the advent of
computers the potential started to be computed in quantum mechanical
terms, first by the Born--Oppenheimer method and, more recently, by
the Car--Parrinello method \cite{carparr} or by path integral
molecular dynamics (see for example \cite{mano}).
\begin{figure}
  \begin{center}
    \includegraphics[width=0.8\textwidth]{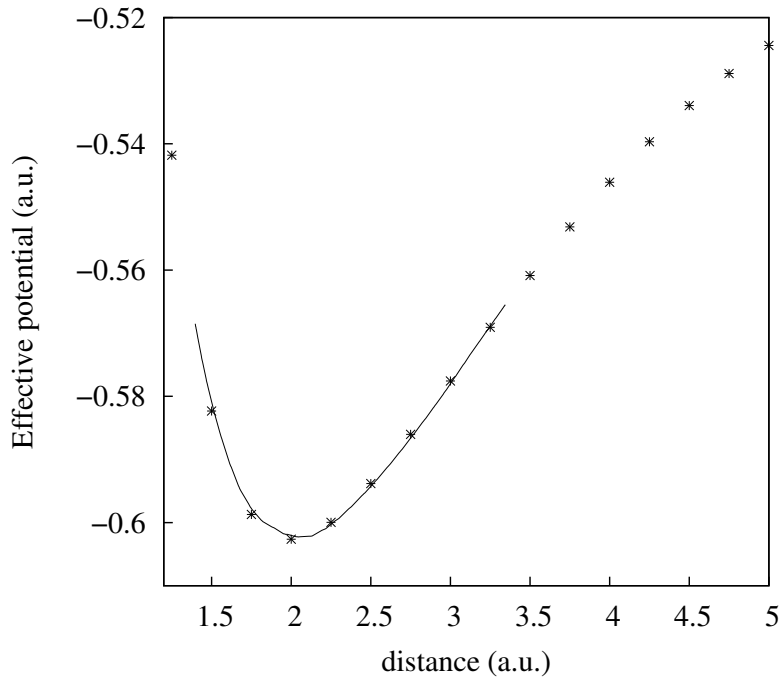}
  \end{center}
  \caption{\label{fig:1} Effective potential as a function of 
    proton distance, computed for suitably chosen initial data
    (continuos line), together with the quantum potential  in
    the Born-Oppenheimer approximation (stars).}
\end{figure}

A natural question is then whether the effective potential acting
among the nuclei can actually be computed in classical terms too,
possibly by molecular dynamics simulations of the Newtonian
trajectories of both ions and electrons. In the present paper we show
that this can be done, at least in the simplest case of the
Hydrogen--molecule ion $H_2^+$, in which the bonding action (between
two protons) is performed by just one electron.  However, while in the
quantum case the electronic state producing the bond can be selected
in a quite natural way (typically as the electronic ground state), the
situation is obvioulsy more complicated in the purely classical case,
since there exists a continuum of electronic states producing the bond
(albeit with relevant constraints, as will be seen later).

The best result we could obtain for the potential by exploring the
electronic initial data is illustrated in Fig.~\ref{fig:1}, where the
potential is seen to superpose in a surprisingly good way the
analogous quantum one, computed in the Born--Oppenheimer approximation
with reference to the ground state, where the semiclassical limit
might be expected not to apply.

\begin{figure}
  \begin{center}
    \includegraphics[width=0.8\textwidth]{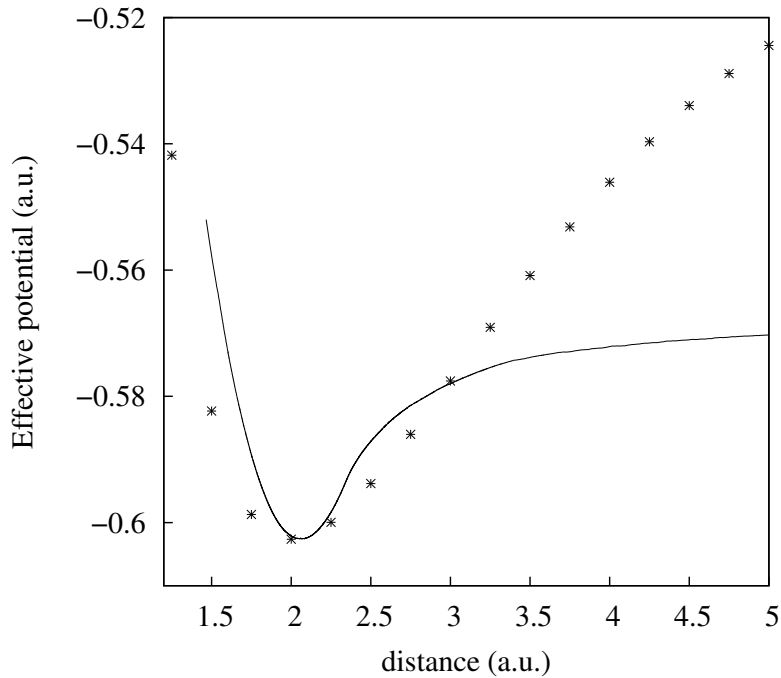}
  \end{center}
  \caption{\label{fig:4} Same as figure~\ref{fig:1} for a ``more 
generic'' choice of the electron initial state.}
\end{figure}

However one should stress that such a result, which is perhaps the
most significant one of the present paper, cannot yet be considered
completely satisfactory in comparison with the quantum one, for at
least two reasons. The first is that the result has a phenomenological
character, since the optimal electronic state was determined not from
first principles, but by a best fit of the experimental data, namely,
the equilibrium distance (bond length) of the protons and their
vibrational frequency. The second reason is that the initial data
producing the result of Fig.~\ref{fig:1} turn out to be somehow
exceptional, as will be discussed later.  However, stable states in
the classical approach can be obtained, if one chooses initial data in
different regions of phase space. An example of the effective
potential obtained in one such region is illustrated in
Fig.~\ref{fig:4}, where the potential is seen to produce a bond, but
to differ very much from the quantum one. Neither are the experimental
data well reproduced.  More details will be given later.

The studies dealing with the possibility of describing the chemical
bond of the $H_2^+$ ion in terms of its Newtonian trajectories have a
long history, with a first phase \cite{bohr, pauli,
  niessen,fermi,langmuir,urey} initiated by the Bohr paper of the year
1913, and a more recent one \cite{turner, strand, muller, howard,
  giapponesi}.  They are performed in the spirit of the ``old quantum
theory'', in which classical trajectories are considered, and
quantization enters only in the choice of the initial data in phase
space.  In all such works the $H_2^+$ ion is modeled as a
nonrelativistic three--body problem with mutual Coulomb interactions
(with appropriate signs). In the approximation of clamped protons the
model reduces to the celebrated two fixed--centers model, which was
established by Euler and Jacobi to be integrable, and thus amenable to
an analytic study.

The model we actually study is an extended version of such a standard
model, inasmuch as the energy of the electron is taken in its
relativistic form. While such a modification might appear to be just a
minor one, it will be shown later to have a relevant impact for the
existence of the effective potential. The reason is that by the
methods of perturbation theory the potential can be proved to exist
only if the system is nearly integrable, whereas the semi-relativistic
model will be shown to present extended chaotic regions. In fact, it
occurs that quasi integrability holds if the angular momentum of the
electron along the inter-nuclear axis is above a theshold of the order
of the reduced Planck constant $\hbar$ (actually, a little smaller
than it).  Thus in the semi-relativistic model the effective potential
exists only for initial data which are dynamically constrained to lie
within a realistic domain. and this makes a consistent fit of the
experimental data possible. Whether this fact be a simple coincidence
or may have a deeper significance, we are unable to say at the moment.

The paper is organized as follows. In Section 2 we illustrate the
three--body model studied here, and also recall the averaging
principle, the pillar of perturbation theory which leads one to
conceive that an effective potential may altogether exist in the
classical case.  In Section 3 we illustrate the method used for the
numerical evaluation of the effective potential, and describe the
results. The conclusions follow in Section 4.

\section{The model and the averaging principle}

We now illustrate, following Born and Heisenberg \cite{bornheis}, how
the possibility itself exists of describing classically the motion of
the protons as decoupled from that of the electron, the only effect of
the latter being of providing an ``effective'' bonding force among the
protons.  The reason is that, in virtue of the great mass difference
between electron and protons, in the full system there exist
``fast''degrees of freedom related to the motion of the electron, and
``slow'' ones related to the protons.  On the other hand, in
perturbation theory the averaging principle states that the system
obtained by averaging over the fast variables describes well (up to a
certain time scale) the motion of the slow ones, on which the system
still depends.  In the standard model of the ion $H_2^+$, i.e.  a
single nonrelativistic electron of mass $m$ interacting with two
(point--like) protons having a much larger mass $M$, all with a charge
of the same modulus $e$, the Hamiltonian is
\begin{equation}\label{totale}
  H= \frac {p^2}{2m} -\frac {e^2}{|\vett r -\vett x_1|}
  -\frac {e^2}{|\vett r -\vett x_2|}
  + \frac{P_1^2}{2M} + \frac{P_2^2}{2M} +\frac { e^2}{|\vett x_1 -\vett
    x_2|}\  ,
\end{equation}
where $\vett p$ and $\vett r$ are the coordinates (momentum and
position) of the electron, while $ \vett P_i$ and $\vett x_i$ are the
coordinates of the two protons.  It is well known that for the
electronic Hamiltonian
\begin{equation} \label{elettrone1}
  H_e= \frac {p^2}{2m} -\frac {e^2}{|\vett r -\vett x_1|}
  -\frac {e^2}{|\vett r -\vett x_2|}
\end{equation}
with  $\vett x_1$ and $\vett x_2$ fixed (the so called Euler two
fixed--centers  problem),
there exist action--angle variables $\vett J, \boldsymbol{\varphi}$ 
such that it  takes the form
\begin{equation} \label{elettrone2}
  H_e= H_e(\vett J, R) \ ,
\end{equation}
which depends only on the actions,
in addition to a parametric
dependence  on the distance
\begin{equation}\label{erre}
R= |\vett x_1-\vett x_2|
\end{equation}
among the protons, i.e. the system is integrable.
\begin{figure}
  \begin{center}
    \includegraphics[width=0.8\textwidth]{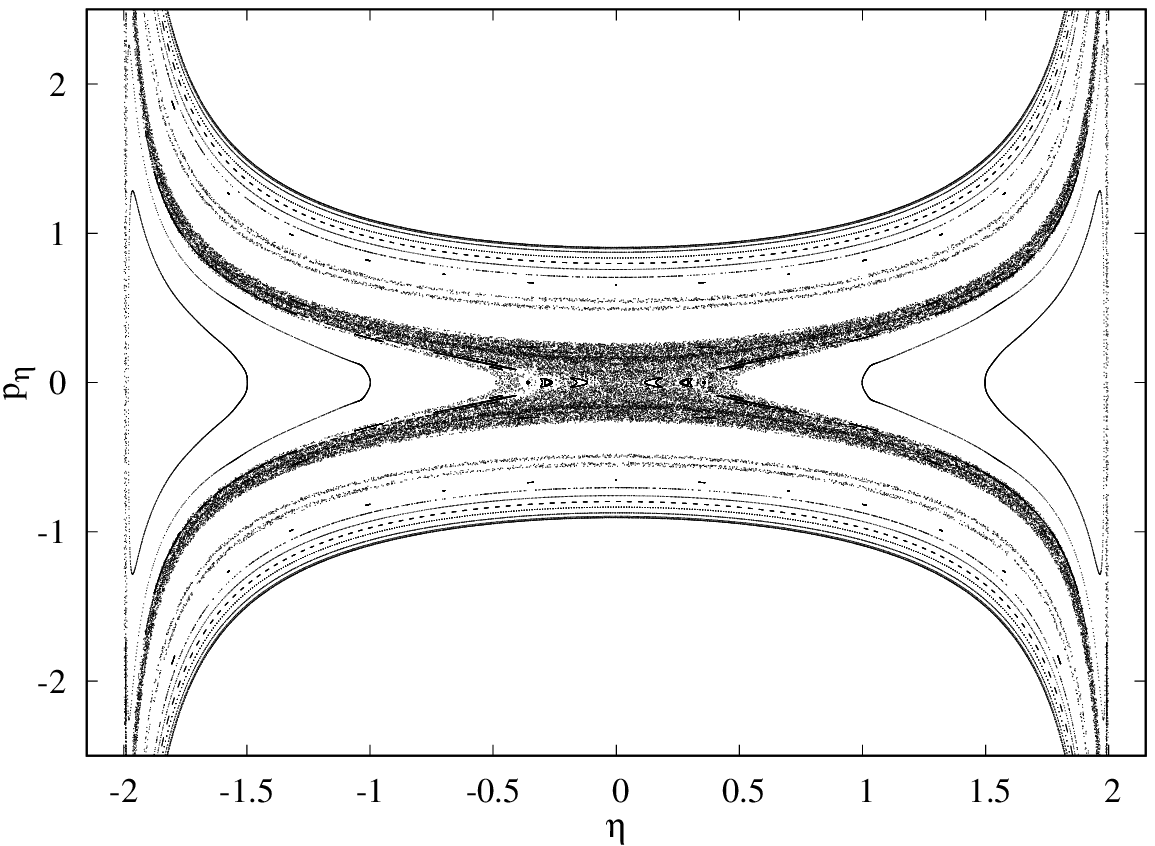}
    \includegraphics[width=0.8\textwidth]{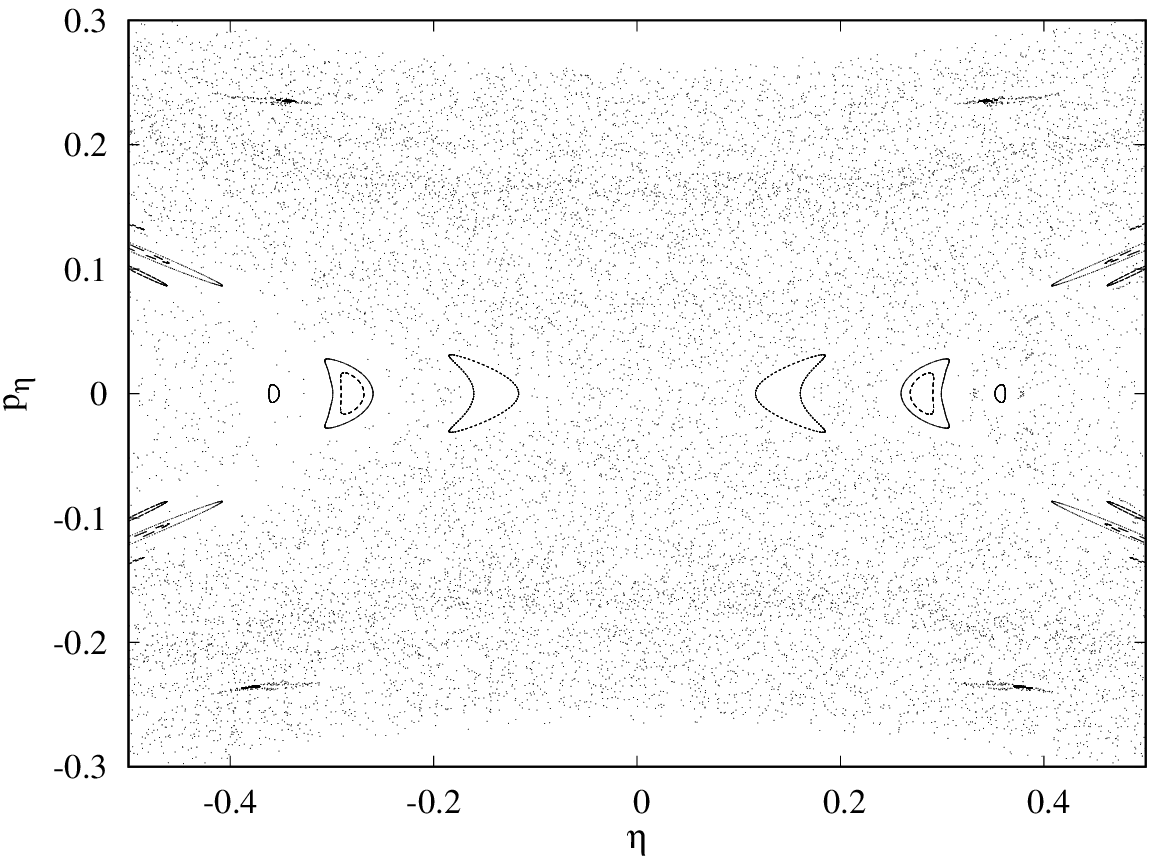}
  \end{center}
  \caption{\label{fig:2} Poincare section $p_\xi = 0$ (see text), for
    the relativistic two fixed--centers model, for energy $E=-0.606$
    and angular momentum $l=0.1$ (in atomic units). Notice that $\eta=0$ corresponds to
    the equatorial plane.  The upper panel shows a large chaotic zone,
    in a region where an orbit should lie in order to be bonding. In the lower
    panel an enlargement of the chaotic zone is shown, exhibiting  some of
    the structures which are present.}
\end{figure}  

Furthermore, the angles $\boldsymbol{ \varphi}$ turn out to be in
general fast variables, i.e.  their frequencies $ \boldsymbol{
  \omega}={\partial H_e}/{\partial \vett J}$ are in general much
larger than the speeds of the other electronic variables.  If now one
applies such a transformation to the full Hamiltonian (\ref{totale}),
in the new variables the Hamiltonian takes the form\footnote{The
  perturbing function $F$ originates from the fact that the canonical
  transformation to the action--angle variables relative to the
  electronic Hamiltonian $H_e$, actually depends parametrically on the
  positions of the protons too, through their distance $R=|\vett
  x_1-\vett x_2|$. So also the generating function $S$ of the global
  canonical transformation depends on $R$, and in particular the
  modulus $|\vett P_1-\vett P_2|$ of relative momentum of the protons
  transforms into $|\vett P_1-\vett P_2|+\partial S/\partial R$.}
\begin{equation}\label{totale2}
  H= H_e(\vett J, R) +
  \frac{P_1^2}{2M} +\frac{P_2^2}{2M} +\frac { e^2}{ R} +
  F (\vett J, \boldsymbol{\varphi}, \vett P_1, \vett P_2,
  R )\ ,
\end{equation}
with a certain function $F$, so that the full Hamiltonian appears as
a ``small'' perturbation of the Hamiltonian
\begin{equation}\label{imperturbata}
 H_0=  \frac{P_1^2}{2M} +\frac{P_2^2}{2M} +\frac { e^2}{R} + H_e (\vett J,R)  \ .
\end{equation}

Now, perturbation theory shows (for a modern development see for example
\cite{vincoli} \cite{giapponesi}) that, if the frequencies
$\boldsymbol{\omega}$ are sufficiently large, then the motion of the
system is ``well'' described by the full Hamiltonian averaged over the
angles, i.e.  essentially by the Hamiltonian $H_0$
(\ref{imperturbata}).  On the other hand, such a Hamiltonian exhibits
in a manifest way the main fact of interest here, namely, that the
electronic energy $H_e (\vett J, R)$ plays the role of an effective
potential among the protons, analogously to what occurs in the quantum
case.  A further study would then establish whether such an effective
potential may overcome the repulsion among the protons, thus ensuring
the existence of a stable state of the ion $H_2^+$.  Actually, one
should rather speak of a ``metastable'' state, because the theorem of
the mean ensures that the result (i.e.  the constancy of the actions
$\vett J$) holds only over a certain time scale, which is long, but
not infinitely long.\footnote{ In order to find really stable states,
  one should use more powerful tools such as KAM theory, which are
  however much more complex. By the way, the theorem of the mean holds
  in an open set of initial data and not for all of them. Namely, it
  holds for the so--called nonresonant set, in which the ratios between the
  frequencies are badly approximated by rational ones.}
\begin{figure}
  \begin{center}
    \includegraphics[width=0.8\textwidth]{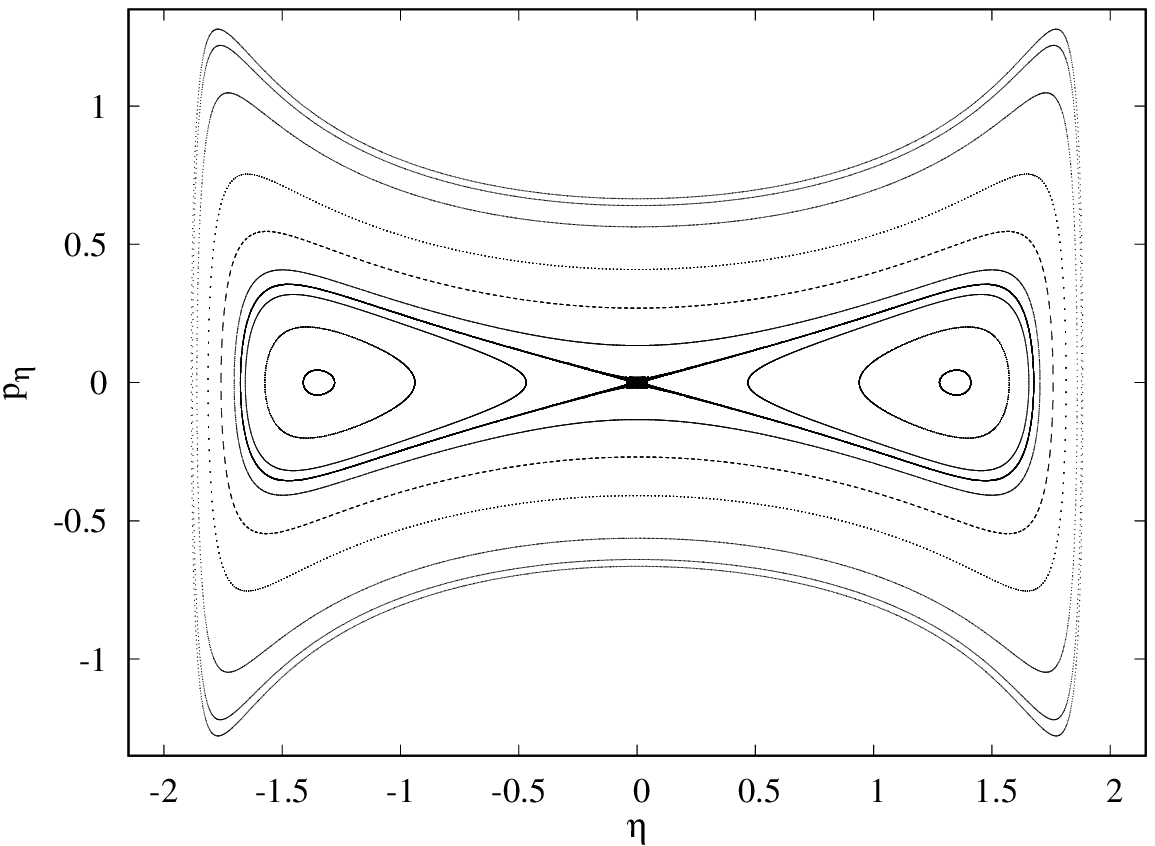}
    \includegraphics[width=0.8\textwidth]{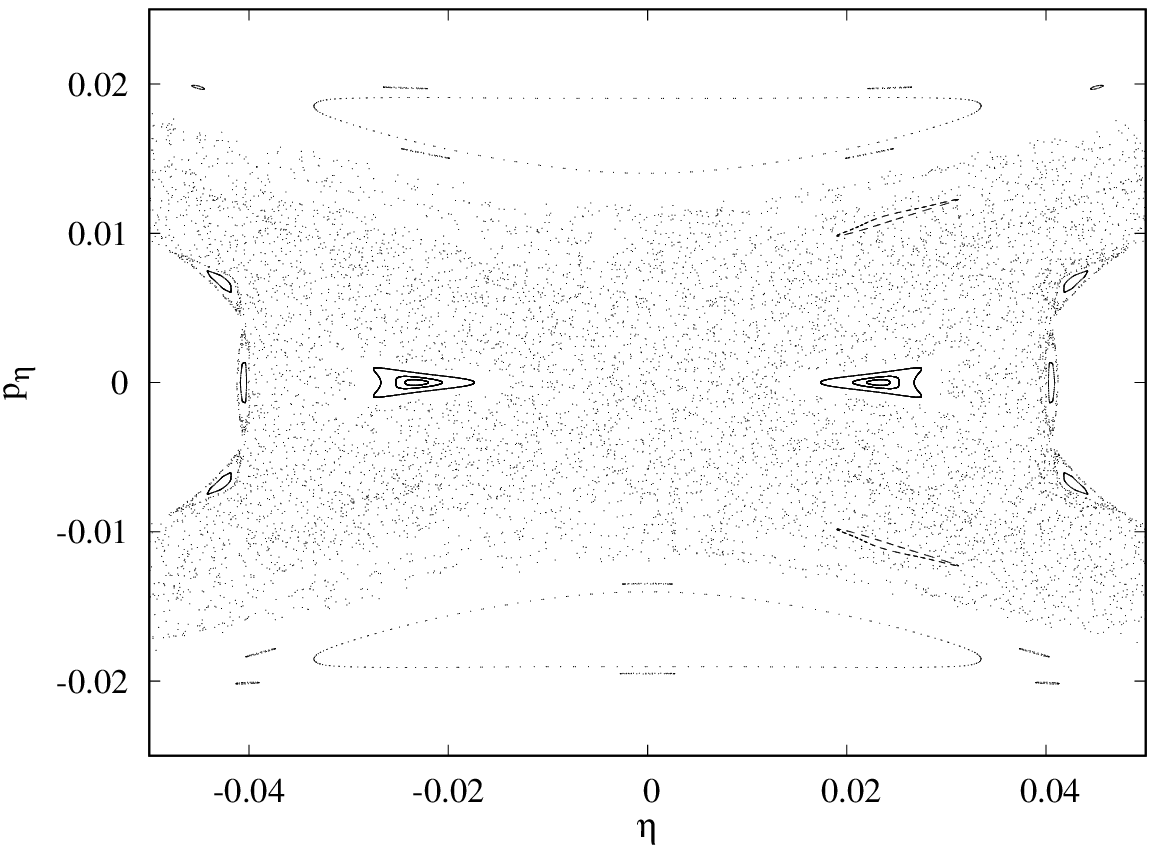}
  \end{center}
  \caption{\label{fig:3} Same as Figure~\ref{fig:2}, still for energy
    $E=-0.606$ but now for a higher value $l=0.6$  of
    the angular momentum. The upper panel shows that now the central
    chaotic zone did shrink, while the vast majority of the orbits
    appear to lie on smooth curves.  In the lower panel an enlargement
    of the chaotic zone is showed, exhibiting some of the structures which
    are present.}
\end{figure}

However, physically the standard model defined by Hamiltonian (\ref{totale}) is
not completely coherent because, for initial data in the atomic
domain, the velocities of the electron quickly become of order of the
speed of light $c$. So we chose to use the partially relativistic
model with Hamiltonian
\begin{equation}\label{totalerel}
  H= mc^2\sqrt{1+ \frac {p^2}{m^2c^2}} -\frac {e^2}{|\vett r -\vett
    x_1|} -\frac {e^2}{|\vett r -\vett x_2|} + \frac{P_1^2}{2M} +
  \frac{P_2^2}{2M} +\frac { e^2}{|\vett x_1 -\vett x_2|}\ .
\end{equation}
But then the electronic energy
\begin{equation} \label{elettrone}
  H_e= mc^2\sqrt{1+ \frac {p^2}{m^2c^2}} -\frac {e^2}{|\vett r -\vett
    x_1|} -\frac {e^2}{|\vett r -\vett x_2|} \ , \quad (\vett x_1,
  \vett x_2 \ \mathrm{fixed}) \ ,
\end{equation}
is no more completely integrable, since it  presents only two
(rather than three) integrals of motion, i.e. the energy and the
component $p_\varphi$ of the angular momentum along the inter-nuclear axis.

The non integrability of the clamped  semi-relativistic 
Hamiltonian (\ref{elettrone}) is
very clearly exhibited through the familiar tool of the ``surfaces of
section'', which we now recall.
Exploiting the constancy of the angular momentum $p_\varphi$, one can
pass to the   corresponding reduced Hamiltonian,  and,  using cylindrical
coordinates
with the $z$ axis along the protons, the electronic Hamiltonian
(\ref{elettrone}) takes the form
\begin{equation}\label{elettronerelcil}
  H_e= mc^2\sqrt{1+  \frac 1{m^2c^2} \big(p_z^2+p_\rho^2+\frac
    {l^2}{\rho^2}\big)}\
  -\frac {e^2} {\sqrt{\rho^2+(z-z_1)^2 }}
  -\frac {e^2}{\sqrt{\rho^2+(z-z_2)^2}}\  .
\end{equation}
where $\rho =\sqrt {x^2+y^2}$ is the  distance of the electron from
the inter-nuclear axis  and $l$ is the value of the angular momentum
$p_\varphi$ of the
electron along that axis. So  one is  now dealing with  a system with two
degrees of freedom in a phase space $\mathbb{R}^4$ and thus, as in the familiar
H\'enon-Heiles case, by fixing the value of energy one is reduced to
a three-dimensional subset (the ``energy surface''). The mapping on
a Poincar\'e surface of section is finally constructed by computing
orbits and intersecting them by a given two-dimensional surface.

In figures~\ref{fig:2} and \ref{fig:3} such a surface is the plane
$p_\xi=0$, where $\xi$ and $\eta$ are the familiar elliptic coordinates
defined (using Arnold's conventions)
by $\xi=|\vett r -\vett x_1|+|\vett r -\vett x_2|$ and
$\eta=|\vett r -\vett x_1|-|\vett r -\vett x_2|$, while $p_\xi$,
$p_\eta$ are the corresponding conjugate momenta.  In Fig~\ref{fig:2},
the values of $\eta$ and $p_\eta$ are reported for $E= 0.606$ and
$l=0.1$, while the distance between the centers is taken equal to 2
(in atomic units).\footnote{We recall that the atomic units 
  are defined by setting $m$=$e$=$\hbar$=1, so that the Bohr radius
  $a_0$ is equal to 1.  The energy of the Hydrogen atom ground state
  turns out to be equal to 0.5, while 0.606 is the energy of the
  electronic ground state for $H_2^+$ ion.}
The whole section is shown in the upper panel, where one sees
that the points corresponding to the different orbits, instead of
being all located on regular curves, as would occur if a third
integral did exist, occupy fuzzy regions, particularly in the central
part. This feature is emphasized in the enlargement of the central
part, which is reported in the lower panel.  A single orbit is seen to
invade a two-dimensional region, and other structures are exhibited,
that one may be tempted to qualify as fractals.
In such a case it is no
more possible to introduce action-angle variables which would make
the Hamiltonian depend on the actions only. Instead, if for the
angular momentum $p_\varphi$ one fixes a larger value such as $l=0.6$,
the surface of section shown in Fig.~\ref{fig:3} appears to be much
more regular, suggesting that in such a case a ``quasi integral of
motion'' exists, the different values of which do identify each of the
invariant curves exhibited.  Such a further integral, by the way,
constitutes in atomic physics the analog of the celebrated ``third
integral'' of celestial mechanics, to which the whole scientific life
of G. Contopoulos was devoted.
In the presence of such a third integral, a
transformation can be found that eliminates the angles from the
electronic Hamiltonian (possibly up to a very small remainder) in a
very extended open set in phase space. In such a situation one might
presume that the full semi-relativistic Hamiltonian (\ref{totalerel})
averaged over the angles provides a good approximation for the motion
of the slow variables, i.e. for the motion of the protons. As
previously explained, in such a situation the electronic energy plays
the role of a potential which complements the repulsive Coulomb
potential among the protons. In the frame of atomic physics, a
discussion presenting some analogies with that given  here, can be
found in the paper \cite{giapponesi}.

\section{Numerical results for the effective potential}

In the previous section it was explained how is it possible at all to
conceive that, analogously to what occurs in quantum mechanics in the
Born-Oppenheimer approximation, in
classical mechanics too the motion of the protons can be described by
eliminating the motion of the electron and replacing it by a suitable
contribution to an effective potential acting among the protons.
More precisely,
this is expected to occur only in
a suitable domain of phase space, where the dynamics of the system is
regular rather than chaotic, i.e. a ``third integral'' exists.
However, the actual implementation of such a program for the full
semi-relativistic Hamiltonian (\ref{totalerel}) considered in this paper,
requires the establishment of delicate results within perturbation theory which, in
view of their complexity, we refrain from explicitly facing here.  By
the way, for the aims indicated in the introduction, such an
investigation would not even be fruitful.
\begin{figure}
  \begin{center}
    \includegraphics[width=0.8\textwidth]{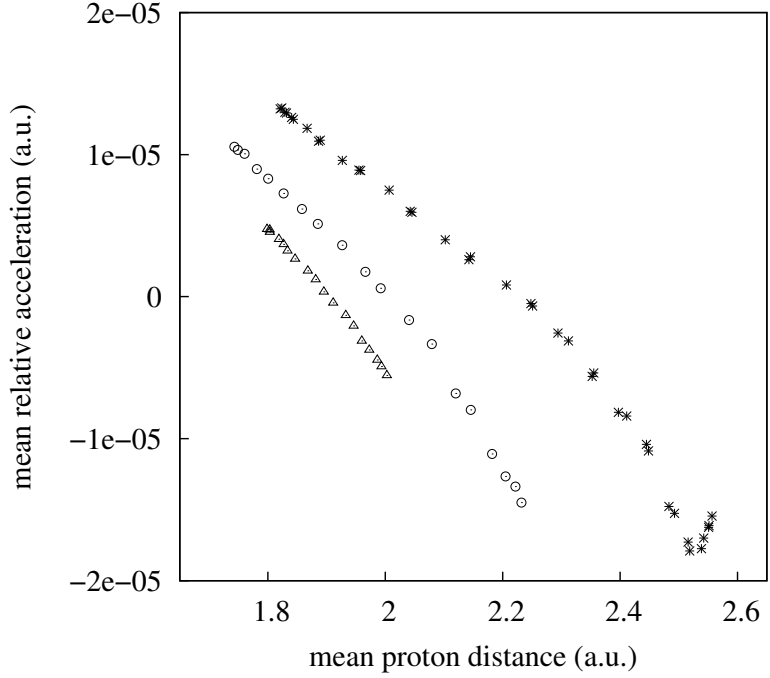}
  \end{center}
  \caption{\label{fig:5}   Radial component of the mean
    relative acceleration  of protons vs. their mean distance,
    for three different
    trajectories. The points are seen to lie on different curves,
    depending on the initial data.}
\end{figure}  

So we  resolved to limit ourselves, in the present work, to
describe a numerical procedure that 
can be used to determine the
effective potential, making reference only to  numerically computed
trajectories.
As shown in the previous section, the effective potential would
emerge if one were able to pass from the actual motion of the electron
to a motion averaged over the associated fast angles.  As such angles
are not well defined in the relativistic case (which we have shown to be
not integrable in the Liouville sense) we decided to replace
such an averaging procedure by a time average over a suitable time-interval
$\Delta t$.\footnote{In our computations we take $\Delta t \simeq 6.4\,10^{-16}$ seconds.} 
So, the equations of motion were numerically integrated,
with a regularized symplectic algorithm that will be described
later,\footnote{The regularization is necessary, since nothing forbids
  the electron from coming arbitrarily close to one of the two protons
  during its motion.}  thus obtaining  trajectories $\vett r(t_k)$,
$\vett x_1(t_k)$, $\vett x_2(t_k)$ of the electron and of the two
protons  respectively.
\begin{figure}
  \begin{center}
    \includegraphics[width=0.8\textwidth]{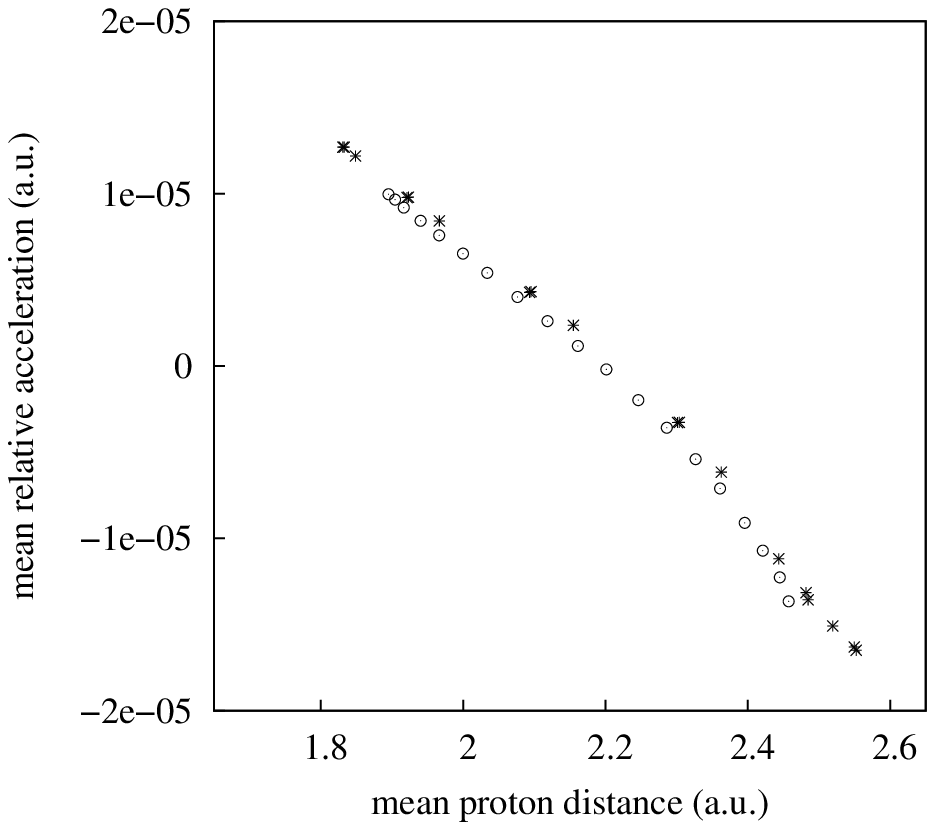}
  \end{center}
  \caption{\label{fig:6}
  Same as Fig.~\ref{fig:5}
  for two different initial data  with the same electronic
  state. The curves now superpose.}
\end{figure}  
Then, time averages were taken of the positions
of the protons, and in particular of their relative distance vector,
namely,
\begin{equation}
  \vett R(t_j) \equal \frac 1{2N} \sum_{k=j-N+1}^{j+N} \big(\vett x_1(t_k) -
  \vett x_2(t_k)\big) \ ,
\end{equation}
where $N$ is determined by the condition $t_{j+N}-t_{j-N+1}=\Delta t$,
whereas the values of $j$ were chosen as multiples of $2N$. The
relative acceleration $\ddot{\vett R}$ at time $t_j$ was then computed
through the usual approximation
\begin{equation}
  \ddot{\vett R} (t_j)\equal \frac{ \vett R(t_{j+1}) + \vett R(t_{j-1}) - 2 \vett
    R(t_j )}{(\Delta t)^2} \ .
\end{equation}
The existence of an effective potential implies that the radial part
$a_R(t_j)$ of the relative acceleration is a function of $R(t_j)$
only, so that, reporting in a graph the pairs
$\big(R(t_j),a_R(t_j)\big)$, the points should be distributed on well
defined curves.  This is exhibited by Figure~\ref{fig:5}, where such
points are reported for three  trajectories, in which the
ion was found to remain stable for the whole integration time,
i.e. for times of the order of picoseconds. The points are seen to lie
on  pretty well defined curves, so that in each case there exists a function
$f(R)$  (depending parametrically on the initial data), such that
\begin{equation}
  a_R = \frac 1\mu f(R)    
\end{equation}
where $\mu$ denotes the reduced mass of the protons. Then, taking a primitive $V(R)$
(with changed sign) of the function $f(R)$, one gets
\begin{equation}
  a_R = -\,  \frac 1\mu \partial_R V(R) \ .    
\end{equation}
Now, the figure shows that the three curves are evidently different,
depending on the chosen initial data.  But this has to be expected
because, according to perturbation theory, the effective potential
depends on the values of the adiabatic invariants (which correspond to
the integrals of motion for the electronic Hamiltonian with clamped
protons).

We thus decided to integrate the equations of motion for several
initial data chosen in a suitable way, i.e. by keeping fixed the
initial value of $R$ and the electronic state, while changing only the
kinetic energy of the protons.  Indeed, in such a way one is assured
that the value of each integral of motion of the electronic system
with clamped protons is the same for all such trajectories.  As one
sees in Fig.~\ref{fig:6}, now the points defined by the pairs
$\big(R(t_j),a_R(t_j)\big)$ are apparently located over a single
pretty well defined curve. Then the potential $V(R)$ can be determined
by integrating numerically, as a function of $R$, the values $\mu a_R$
found: actually this obviously determines the potential, up to an
additive constant.

A typical form of the effective potential thus found is exhibited in
Fig.~\ref{fig:4}. The initial data for the electron were chosen as
follows: the energy $E$ was fixed at the experimental value, while the
value $l$ of the component of the angular momentum $p_\varphi$ along
the inter-nuclear axis was set equal to $0.6$.  In this way one is
assured that the electronic Hamiltonian with clamped potential, as
shown in Figure~\ref{fig:3}, is essentially integrable. Then we find
that there exists a value for the ``third integral'' such that the
equilibrium distance is equal to the experimental one.  The additive
constant, which in principle could be fixed by imposing $V\to 0$ as
$R\to \infty$,\footnote{This however cannot be accomplished by
  numerical computations.} was instead chosen in such a way that the
minimum of the effective potential coincide with the minimum of the
Born-Oppenheimer quantum potential.

From the qualitative point of view the result might be considered
satisfactory, since it exhibits that a bonding effect exists in a
classical frame.  Quantitatively, however, the result is not so good,
because not only the quantum potential is not well reproduced, but
also the vibrational frequency is found to be about one and a half
times larger than the experimental one. One should thus perform a
systematic exploration of the possible electronic states, in order to
check whether a better agreement with the experimental data can be
found, which we didn't do. We only observed that the result just
illustrated is the best one in a neighbourhood of the particular state
considered, because larger values of the oscillation frequency were
always found.

However, following an old suggestion advanced by
Langmuir\cite{langmuir} and particularly by Urey\cite{urey}, it
occured to us to find that there exist electronic states in a
different region which lead to results that are apparently better, the
best of which is reported in Figure~\ref{fig:1}. We considered in fact
electronic motions in the equatorial plane (i.e. in the plane of
symmetry for the ions, normal to the internuclear axis), in particular
near to motions taking place in a straight line perpendicular to and
passing through the internuclear axis.  The electronic state was
chosen in order fit the experimental values of the bond length and of
the vibrational frequency of the protons.\footnote{In atomic units,
  the electron energy was taken equal to $E=-0.69$ while the angular
  momentum was fixed to $l=0.97$. For what concerns the initial data
  for the protons, they were taken at a distance equal to the bond
  length, with opposite speeds.}  i.e. the quadratic part of the
potential.  Startingly, the classical effective potential is seen to
actually fit the quantum one not only in the linear regime, but also
in an extended nonlinear one.  However, while in the previous phase
space region the ion was stable with respect to changes of the initial
data in an open domain, in the latter case the ion is stable only if
the protons have initial velocities along the internuclear axis.
Otherwise the ion splits into a proton and an Hydrogen atom.

We finally end this section with a short description of the
integration method, which is indeed standard in stellar dynamics
simulations, and we actually took from the paper \cite{mikkola}.  As
was already pointed out, during its motion the electron can come very
close to any of the two protons and thus, in order to keep the
precision of the numerical integration, the integration step has to be
reduced.  But this is likely to prejudice the symplectic character of
the integration algorithm. To avoid this, in the work \cite{mikkola}
it was proposed to regularize the equations of motion by using, in
place of the time $t$, the variable $s$ defined by
\begin{equation}
  \dif s \equal \frac {\dif t}{U} \ ,
\end{equation}
$U$ being the potential energy of the system.  In such a way it turns
out that to equal increments of the variable $s$ correspond time
increments which are very small near the singular points of the
potential, where it diverges. After the change of variable the
equations of motion preserve the Hamiltonian form, with the only
difference that, instead of the original Hamiltonian $H=T+U$, where
$T$ and $U$ are the kinetic and the potential energies, the
Hamiltonian now takes the form
\begin{equation}
  H' \equal \log(T-E) + \log(-U)  \ ,
\end{equation}
where $E$ is the value of $H$ determined by the initial data.  The
only difference is that for the kinetic energy $T$ of the electron we
used the relativistic formula; moreover in $U$ there appears a
repulsive part which obviously doesn't show up in the case of stellar
dynamics.\footnote{ However one easily checks that, if the total
  energy $E$ is negative, the potential $U$ remains negative, and thus
  the Hamiltonian $H'$ turns out to be well defined.}  The equations
of motion were integrated using the leap-frog algorithm (which is well
known to be symplectic), whereas $t$ was obtained by computing the
definite integral $\int_0^t U \dif s$ through the trapezoidal rule.

\section{Conclusions}
It seems to us that the most significant  result  emerging from the
present study is   that, at least in the simplest
possible  case of systems  involving just one electron,
a relevant physical quantity such 
as the effective potential which bonds a molecule,
can be   computed in terms
of  Newtonian electronic  trajectories. 
This is somehow  disconcerting, since  the result  was obtained
for states  in which the de Broglie wavelength of the electron
is of the order of the
dimensions of the molecule,  so that, according to the
accepted wisdom,  the electronic trajectories would appear to make no  physical
sense. This seems to support the idea, particularly
pursued  by Gutzwiller \cite{gutz} (see also \cite{amico,giap2}),
that the original conception 
of the correspondence principle should be extended to some more general one,
which was not yet formulated.

A further relevant  fact seems to be  the introduction of the
relativistic correction, which  apparently should  be    necessary in atomic
physics   if one  decides to take electronic trajectories into
consideration.
Indeed such a fact implies  the nonintegrability of the
reference unperturbed  model, and thus  the occurring of chaoticity regions
in which the adiabatic principle does  not apply. So the potential exists
only in a  definite region of phase space, which, by the way, entails that
the relevant action defining the electronic initial data should be larger
than a value of the order of the reduced Planck constant $\hbar$

Another relevant issue concerns, more concretely, the program of
extending molecular dynamics techniques from
the motions of the ions to those  of the electrons. Here, however,
one meets with
 a severe difficulty. Indeed, already in the case  studied
 in the present paper  which involves only one electron,
 the trajectories  that   best reproduce the experiments
 (bond length and vibrational frequency)  and the quantum  potential, 
turn out to be  unstable under small perturbations, for example perturbations
of the initial data. On the other hand,  some form  of
instability  seems to be   quite a general
feature plaguing  classical models of atomic physics. Indeed, 
since the time of Nicholson\cite{nich} and Bohr\cite{bohr}
analogous instabilities
were always observed  for  atomic  systems
involving  more than just one electron.
To cope with  such an instability problem seems to be the priority,
if one aims at effectively
extending  to electrons the methods of molecular dynamics,  and more in
general at stressing  the significance of Newtonian  electronic
trajectories in atomic physics. The impressively  good reproduction of
the quantum potential  exhibited in Figure~\ref{fig:1}, which can
hardly be regarded as  fortuitous, seems to indicate that some progress
may hopefully be accomplished  along these lines.

\end{document}